\documentclass[journal]{IEEEtran}

\usepackage{cite}
\usepackage{amsmath,amssymb,amsfonts}
\usepackage{algorithmic}
\usepackage{graphicx}
\usepackage{textcomp}
\usepackage{xcolor}
\usepackage{multirow}
\usepackage{float}

\usepackage[font=scriptsize]{subcaption}
\usepackage[font=footnotesize]{caption}
\usepackage[acronyms,nonumberlist,nopostdot,nomain,nogroupskip]{glossaries}

\usepackage{booktabs}

\usepackage{tikz}
\usepackage{pgfplots}
\pgfplotsset{compat=newest} 
\pgfplotsset{plot coordinates/math parser=false} 
\newlength\fheight
\newlength\fwidth
\usetikzlibrary{plotmarks,patterns,decorations.pathreplacing,backgrounds,calc,arrows,arrows.meta,spy,matrix}
\usepgfplotslibrary{patchplots,groupplots}
\usepackage{tikzscale}
\usepackage{siunitx}

\usepackage{multirow}
\usepackage{tkz-kiviat}
\usepackage{tikz-qtree}
\usetikzlibrary{trees} 
\usepackage{siunitx}
\usetikzlibrary{fadings}

\tikzfading[name=middle,
            top color=transparent!100,
            bottom color=transparent!100,
            middle color=transparent!20]

\usetikzlibrary{arrows,automata,calc,shapes, positioning,shadows,shadows.blur,shapes.geometric}

\newacronym{3gpp}{3GPP}{3rd Generation Partnership Project}
\newacronym{itu}{ITU}{International Telecommunication Union}
\newacronym{adc}{ADC}{Analog to Digital Converter}
\newacronym{5g}{5G}{5th generation}
\newacronym{6g}{6G}{6th generation}
\newacronym{aimd}{AIMD}{Additive Increase Multiplicative Decrease}
\newacronym{am}{AM}{Acknowledged Mode}
\newacronym{amc}{AMC}{Adaptive Modulation and Coding}
\newacronym{aqm}{AQM}{Active Queue Management}
\newacronym{awgn}{AGWN}{Additive White Gaussian Noise}
\newacronym{balia}{BALIA}{Balanced Link Adaptation}
\newacronym{bdp}{BDP}{Bandwidth-Delay Product}
\newacronym{bf}{BF}{beamforming}
\newacronym{cc}{CC}{Congestion Control}
\newacronym{cdf}{CDF}{Cumulative Distribution Function}
\newacronym{cn}{CN}{Core Network}
\newacronym{cqi}{CQI}{Channel Quality Information}
\newacronym{cp}{CP}{Control Plane}
\newacronym{csirs}{CSI-RS}{Channel State Information - Reference Signal}
\newacronym{dc}{DC}{Dual Connectivity}
\newacronym{rb}{RB}{Resource Block}
\newacronym{dce}{DCE}{Direct Code Execution}
\newacronym{dci}{DCI}{Downlink Control Information}
\newacronym{udp}{UDP}{User Datagram Protocol}
\newacronym{dl}{DL}{Downlink}
\newacronym{dmr}{DMR}{Deadline Miss Ratio}
\newacronym{dmrs}{DMRS}{DeModulation Reference Signal}
\newacronym{e2e}{E2E}{End-to-End}
\newacronym{ppp}{PPP}{Poission Point Process}
\newacronym{si}{SI}{Study Item}
\newacronym{ecn}{ECN}{Explicit Congestion Notification}
\newacronym{edf}{EDF}{Earliest Deadline First}
\newacronym{enb}{eNB}{eNodeB}
\newacronym{epc}{EPC}{Evolved Packet Core}
\newacronym{es}{ES}{Edge Server}
\newacronym{cav}{CAV}{Connected and Autonomous Vehicle}
\newacronym{fdma}{FDMA}{Frequency Division Multiple Access}
\newacronym{fdd}{FDD}{Frequency Division Duplexing}
\newacronym{upa}{UPA}{Uniform Planar Array}
\newacronym[firstplural=Radio Access Technologies (RATs)]{rat}{RAT}{Radio Access Technology}
\newacronym[firstplural=Radio Access Technology (RTs)]{rt}{RT}{Radio Technology}
\newacronym{fs}{FS}{Fast Switching}
\newacronym{isd}{ISD}{inter-site distance}
\newacronym{ftp}{FTP}{File Transfer Protocol}
\newacronym{gnb}{gNB}{Next Generation Node Base}
\newacronym{harq}{HARQ}{Hybrid Automatic Repeat reQuest}
\newacronym{hetnet}{HetNet}{Heterogeneous Network}
\newacronym{hh}{HH}{Hard Handover}
\newacronym{hol}{HOL}{Head-of-Line}
\newacronym{ia}{IA}{Initial Access}
\newacronym{imt}{IMT}{International Mobile Telecommunication}
\newacronym{iot}{IoT}{Internet of Things}
\newacronym{los}{LOS}{Line of Sight}
\newacronym{lte}{LTE}{Long Term Evolution}
\newacronym{m2m}{M2M}{Machine to Machine}
\newacronym{mac}{MAC}{Medium Access Control}
\newacronym{mc}{MC}{Multi-Connectivity}
\newacronym{mcs}{MCS}{Modulation and Coding Scheme}
\newacronym{mec}{MEC}{Mobile Edge Cloud}
\newacronym{mi}{MI}{Mutual Information}
\newacronym{mimo}{MIMO}{Multiple Input Multiple Output}
\newacronym{mmwave}{mmWave}{millimeter wave}
\newacronym{mptcp}{MPTCP}{Multipath TCP}
\newacronym{mr}{MR}{Maximum Rate}
\newacronym{mss}{MSS}{Maximum Segment Size}
\newacronym{mtd}{MTD}{Machine-Type Device}
\newacronym{mtu}{MTU}{Maximum Transmission Unit}
\newacronym{nfv}{NFV}{Network Function Virtualization}
\newacronym{vnf}{VNF}{ Virtualization Network Function}
\newacronym{sdn}{SDN}{Software Defined Networking}
\newacronym{nlos}{NLOS}{Non Line of Sight}
\newacronym{nlosb}{NLOSb}{Building Non Line of Sight}
\newacronym{nlosv}{NLOSv}{Vehicle Non Line of Sight}
\newacronym{nr}{NR}{New Radio}
\newacronym{ofdm}{OFDM}{Orthogonal Frequency Division Multiplexing}
\newacronym{pdcch}{PDCCH}{Physical Downlonk Control Channel}
\newacronym{pdcp}{PDCP}{Packet Data Convergence Protocol}
\newacronym{pdsch}{PDSCH}{Physical Downlink Shared Channel}
\newacronym{pdu}{PDU}{Packet Data Unit}
\newacronym{pf}{PF}{Proportional Fair}
\newacronym{pgw}{PGW}{Packet Gateway}
\newacronym{phy}{PHY}{Physical}
\newacronym{pbch}{PBCH}{Physical Broadcast Channel}
\newacronym[plural=\gls{mme}s,firstplural=Mobility Management Entities (MMEs)]{mme}{MME}{Mobility Management Entity}
\newacronym{prb}{PRB}{Physical Resource Block}
\newacronym{pss}{PSS}{Primary Synchronization Signal}
\newacronym{pucch}{PUCCH}{Physical Uplink Control Channel}
\newacronym{pusch}{PUSCH}{Physical Uplink Shared Channel}
\newacronym{rach}{RACH}{Random Access Channel}
\newacronym{ran}{RAN}{Radio Access Network}
\newacronym{red}{RED}{Random Early Detection}
\newacronym{rf}{RF}{Radio Frequency}
\newacronym{rlc}{RLC}{Radio Link Control}
\newacronym{rlf}{RLF}{Radio Link Failure}
\newacronym{rrc}{RRC}{Radio Resource Control}
\newacronym{rrm}{RRM}{Radio Resource Management}
\newacronym{rr}{RR}{Round Robin}
\newacronym{rs}{RS}{Remote Server}
\newacronym{rsrp}{RSRP}{Reference Signal Received Power}
\newacronym{rss}{RSS}{Received Signal Strength}
\newacronym{rtt}{RTT}{Round Trip Time}
\newacronym{rw}{RW}{Receive Window}
\newacronym{rx}{RX}{Receiver}
\newacronym{sa}{SA}{standalone}
\newacronym{sack}{SACK}{Selective Acknowledgment}
\newacronym{sap}{SAP}{Service Access Point}
\newacronym{sch}{SCH}{Secondary Cell Handover}
\newacronym{scoot}{SCOOT}{Split Cycle Offset Optimization Technique}
\newacronym{sdma}{SDMA}{Spatial Division Multiple Access}
\newacronym{sinr}{SINR}{Signal to Interference plus Noise Ratio}
\newacronym{sm}{SM}{Saturation Mode}
\newacronym{snr}{SNR}{Signal to Noise Ratio}
\newacronym{son}{SON}{Self-Organizing Network}
\newacronym{ss}{SS}{Synchronization Signal}
\newacronym{srs}{SRS}{Sounding Reference Signal}
\newacronym{sss}{SSS}{Secondary Synchronization Signal}
\newacronym{tb}{TB}{Transport Block}
\newacronym{tcp}{TCP}{Transmission Control Protocol}
\newacronym{tdd}{TDD}{Time Division Duplexing}
\newacronym{tdma}{TDMA}{Time Division Multiple Access}
\newacronym{tfl}{TfL}{Transport for London}
\newacronym{tm}{TM}{Transparent Mode}
\newacronym{prr}{PRR}{Packet Reception Ratio}
\newacronym{trp}{TRP}{Transmitter Receiver Pair}
\newacronym{tti}{TTI}{Transmission Time Interval}
\newacronym{ttt}{TTT}{Time-to-Trigger}
\newacronym{tx}{TX}{Transmitter}
\newacronym{ue}{UE}{User Equipment}
\newacronym{ul}{UL}{Uplink}
\newacronym{uml}{UML}{Unified Modeling Language}
\newacronym{um}{UM}{Unacknowledged Mode}
\newacronym{utc}{UTC}{Urban Traffic Control}
\newacronym{vm}{VM}{Virtual Machine}
\newacronym{rsrq}{RSRQ}{Reference Signal Received Quality}
\newacronym{rssi}{RSSI}{Received Signal Strength Indicator}
\newacronym{crs}{CRS}{Cell Reference Signal}
\newacronym{v2v}{V2V}{Vehicle-to-Vehicle}
\newacronym{v2i}{V2I}{Vehicle-to-Infrastructure}
\newacronym{v2n}{V2N}{Vehicle-to-Network}
\newacronym{v2x}{V2X}{Vehicle-to-Everything}
\newacronym{vn}{VN}{Vehicular Node}
\newacronym{dsrc}{DSRC}{Dedicated Short Range Communication}
\newacronym{ci}{CI}{context information}
\newacronym{voi}{VoI}{value of information}
\newacronym{gps}{GPS}{Global Positioning System}
\newacronym{qos}{QoS}{Quality of Service}
\newacronym{qoe}{QoE}{Quality of Experience}
\newacronym{ml}{ML}{Machine Learning}
\newacronym{ahp}{AHP}{Analytic Hierarchy Process}
\newacronym{lidar}{LIDAR}{Light Detection and Ranging}
\newacronym{sumo}{SUMO}{Simulation of Urban MObility}
\newacronym{wave}{WAVE}{Wireless Access in Vehicular Environment}
\newacronym{c-its}{C-ITS}{Connected Intelligent Transportation System}
\newacronym{dash}{DASH}{Dynamic Adaptive Streaming over HTTP}
\newacronym{http}{HTTP}{HyperText Transfer Protocol}
\newacronym{nt}{NT}{non-terrestrial}
\newacronym{ntc}{NTC}{non-terrestrial communication}
\newacronym{ntn}{NTN}{non-terrestrial network}
\newacronym{haps}{HAPS}{High Altitude Platform Station}
\newacronym{hap}{HAP}{High Altitude Platform}
\newacronym{lap}{LAP}{Low Altitude Platform}
\newacronym{leo}{LEO}{Low Earth Orbit}
\newacronym{meo}{MEO}{Medium Earth Orbit}
\newacronym{geo}{GEO}{Geostationary Earth Orbit}
\newacronym{uav}{UAV}{Unmanned Aerial Vehicle}
\newacronym{nsat}{nSAT}{Nanosatellite}
\newacronym{ehf}{EHF}{extremely high-frequency}
\newacronym{ioe}{IoE}{Internet of Everyone}
\newacronym{gan}{GaN}{Gallium Nitride}

\def\BibTeX{{\rm B\kern-.05em{\sc i\kern-.025em b}\kern-.08em
    T\kern-.1667em\lower.7ex\hbox{E}\kern-.125emX}}
\begin{document}

\title{The Potential of Multilayered Hierarchical Nonterrestrial Networks for 6G: A Comparative Analysis Among Networking Architectures}
\author{{Dengke Wang,~\IEEEmembership{Student Member, IEEE}, Marco Giordani,~\IEEEmembership{Member, IEEE},\\ Mohamed-Slim Alouini,~\IEEEmembership{Fellow, IEEE},
and Michele Zorzi,~\IEEEmembership{Fellow, IEEE}}

\thanks{
D. Wang and M.-S. Alouini are with the King Abdullah University of Science and Technology (KAUST), Thuwal, Saudi Arabia (email: \{dengke.wang,slim.alouini\}@kaust.edu.sa).
M. Giordani (\emph{corresponding author}) and M. Zorzi are with the Department of Information Engineering (DEI), University of Padova, Italy  (email: \{giordani,zorzi\}@dei.unipd.it).
}
}

\maketitle

\begin{abstract}
6th generation (6G) communication research is currently focusing on non-terrestrial networks (NTNs) to promote ubiquitous and ultra-high-capacity global connectivity. Specifically, multi-layered hierarchical networks, i.e., the orchestration among different aerial/space platforms, including Low and High Altitude Platforms (LAPs and HAPs), and satellites co-operating at different altitudes, currently represent one the most attractive technological options to solve coverage and latency constraints associated with the NTN paradigm. However, there are still several issues to be resolved for proper network design. 
In this work, we evaluate the performance of different multi-layered non-terrestrial configurations, and provide guidelines on the optimal working point(s) for which it is possible to achieve a good compromise between improved system flexibility and network performance, with respect to a baseline standalone deployment. 
\end{abstract}

      \begin{tikzpicture}[remember picture,overlay]
\node[anchor=north,yshift=-10pt] at (current page.north) {\parbox{\dimexpr\textwidth-\fboxsep-\fboxrule\relax}{
\centering\footnotesize This paper has been accepted for publication in IEEE Vehicular Technology Magazine, \textcopyright 2021 IEEE.\\
Please cite it as: D. Wang, M. Giordani, M.-S. Alouini, M. Zorzi, “The Potential of Multilayered Hierarchical Nonterrestrial Networks for 6G: A Comparative Analysis Among Networking Architectures”, IEEE Vehicular Technology Magazine, 2021.}};
\end{tikzpicture}

\section{Introduction}
With  \gls{5g} wireless networks ready for commercial roll-out, \gls{6g} research~\cite{giordani2020toward} is currently concentrating on the development of \glspl{ntn} in which \glspl{lap}, \glspl{hap}, and satellites expand traditional bi-dimensional network designs to operate in the three-dimensional space~\cite{giordani2019non}.
In particular, air/spaceborne stations can assist terrestrial infrastructures in promoting flexible global connectivity in crowded areas,  cost-effective network coverage in public safety situations, and last-mile service delivery and backhaul in remote/rural/hard-to-access zones~\cite{chaoub20206g}.
In this perspective, the  \gls{3gpp} has  approved the first Rel-17 specifications to support NTNs in 5G NR systems, and study items are  encouraged for Rel-18 and Rel-19, thus acknowledging long-term research towards~6G~\cite{38821}.

Until the advent of~\gls{5g}, \glspl{ntn} have been mainly relegated to support services like television broadcasting, meteorology, and navigation, which are generally provided by standalone satellite constellations operating on \glspl{leo} or \glspl{geo}.
On one side, satellites offer a number of advantages to ground users, including extremely large coverage regions and favorable line-of-sight connectivity.
On the other side, satellite deployments typically suffer from severe path loss, huge communication delays, and expensive installation costs~\cite{cianca2005integrated}.  
These weaknesses can be efficiently downplayed by complementing satellite systems with the participation of \glspl{hap} and \glspl{lap}. 
In addition to their versatility and lower cost, these elements achieve lower latency thanks to the shorter distances involved, and can operate with spot beams delivering more capacity to ground users, more  diversity options, or a combination of the two.
However, not only do LAPs, specifically \glspl{uav}, incur significant energy consumption for propulsion and hovering, but also the intrinsic mobility of HAPs could impair network performance without proper coordination and countermeasures. 
In this context, interconnecting space, air, and ground networks is emerging as a viable approach to enhance communication, with each network segment compensating for the weaknesses of the other.

Initial studies have demonstrated that the availability of multi-layered hierarchical networks can provide better coverage, resilience, and flexibility compared to standalone deployments, which makes them suitable for several practical fields in future networks, including  traffic control and emergency communication~\cite{liu2018space}. 
Software-defined networks may also offer a programmable, scalable, and customizable framework to integrate space, air, and ground components for matching traffic demands with the available network capacity~\cite{zhou2018sagecell}.
In recent years, private organizations have also financed projects to provide broadband internet to the world by combining the persistence of satellites and HAPs with the flexibility of UAVs, such as Airbus Zephyr's initiative.
However, unlike traditional standalone architectures, multi-layered NTNs are affected by limitations related to traffic distribution, resource allocation, load balancing, and mobility management, which require \emph{end-to-end} (rather than \emph{point-to-point}) optimization. In this context, there are still various questions to be answered for proper network design, in particular which degree of integration results in better spatio-temporal coverage.



\begin{figure*}[t!]
\centering 
\includegraphics[width=0.99\textwidth]{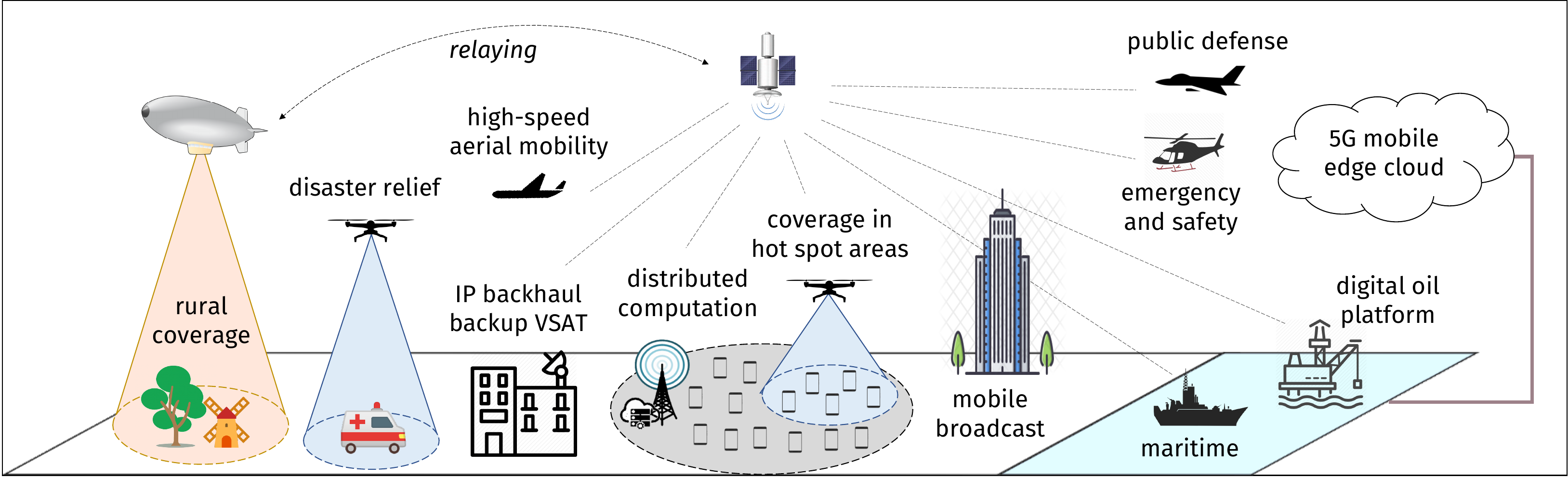}
\caption{Use cases enabled by non-terrestrial networks.}
\label{fig:usecase}
\end{figure*}

Along these lines, in this work we evaluate the performance of different hierarchical NTN architectures in various scenarios, and provide guidelines on the working point(s) that achieve a good compromise between system flexibility and network capacity, against a baseline standalone GEO deployment.
Unlike prior works, which typically consider hybrid satellite terrestrial networks only, we study whether a fully integrated space-air-ground architecture may result in better coverage. Moreover, we characterize which combinations of parameters, including the operational frequency, the deployment altitude, the antenna configuration, and the quality of service requirements, result in optimized network performance.
To this aim, we adopt the architectural designs and related system parameters provided by the \gls{3gpp} and the \gls{itu}  technical reports for NTNs, thus guaranteeing accurate and realistic system-level analyses. 
Our results demonstrate that, while GEO satellite operations can be improved by deploying intermediate stratospheric HAPs, power/antenna constraints on spaceborne vehicles make it undesirable to adopt LEO satellites to relay the upstream GEO signals. 

The remainder of this paper is organized as follows. 
In Sec.~\ref{sec:multi} we discuss the main NTN use cases and possible methods to cooperatively integrate non-terrestrial layers, in Sec.~\ref{sec:design} we describe our system model, and in Sec.~\ref{sec:results} we present our main findings and simulation results. Concluding remarks are finally provided in Sec.~\ref{sec:conclusions}.

\section{Multi-Layered Non-Terrestrial Networks}
\label{sec:multi}

NTNs play a leading role in 5G and beyond by covering different verticals, including health care, intelligent transportation, public safety, and many others. 
In these regards, multi-layered NTNs can further exploit the complementary advantages of space, air, and ground facilities to make the best use of the 3D paradigm. In Sec.~\ref{sub:ntn_use_cases}, we review potential NTN use cases, while in Sec.~\ref{sub:integrating_non_terrestrial_layers}, we present possible solutions to integrate non-terrestrial layers.

\subsection{NTN Use Cases} 
\label{sub:ntn_use_cases}

Recent technological innovations in the aerial/space industry have made it possible to enable  advanced use cases for NTNs~\cite{rinaldi2020non,giordani2020satellite}, as illustrated in Fig.~\ref{fig:usecase} and discussed below.

\begin{itemize}
    \item \textit{Service Continuity}, i.e., providing connectivity when terrestrial networks are overloaded (e.g., in hot spot areas or during rush hours) or in those regions where installation of terrestrial infrastructure is too expensive or even impossible (e.g., above oceans or deserts). NTNs can also improve communication availability in remote or disaster zones, or to preserve the connection in emergency situations when the primary terrestrial path is out of service. For these types of services, reliability is a critical requirement: the interruption time for data packet exchange should be maintained below 1 ms.
    \item \textit{Service Ubiquity}, i.e., guaranteeing global umbrella coverage via aerial/space links.  The wide broadcast nature of non-terrestrial platforms can indeed support multimedia content provisioning to wide geographical areas, and serve wireless backhaul traffic requests in rural locations with no fiber backhaul solutions. In the next 10 years, NTNs are expected to provide a low-latency connection of at least 10 Mbps in rural/remote areas, and a 75\% Internet penetration worldwide.
    \item \textit{Service Scalability}, i.e., offloading traffic from terrestrial networks to more computationally powerful space/air nodes for timely (and aggregated) processing of data, while allowing simplified hardware design on the ground. NTNs also promote energy sustainability as aerial platforms, operated through renewable sources, can be deployed on demand, unlike always-on terrestrial~stations.
\end{itemize}

\begin{figure*}[t!]
\centering 
\includegraphics[width=0.99\textwidth]{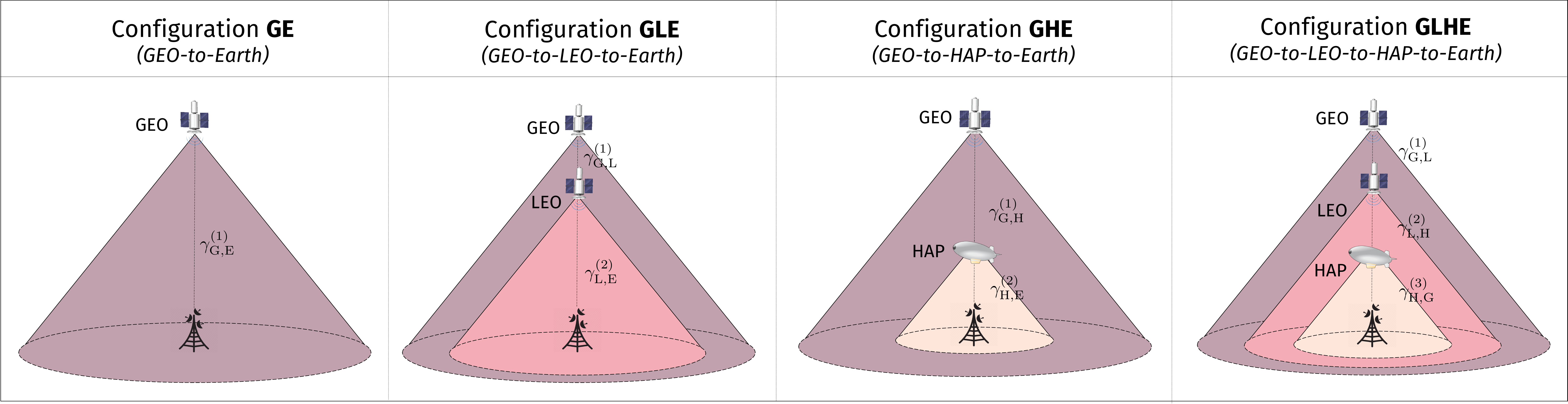}
\caption{Possible multi-layered  NTN configurations based on a GEO satellite. In Configuration GE direct GEO-ground communication is involved; in Configuration GLE (GHE) a LEO (HAP) bridges the GEO communications towards the ground; in Configuration GLHE a complete integrated space-air-ground architecture is considered.}
\label{fig:1}
\end{figure*}

\subsection{Integrating Non-Terrestrial Layers} 
\label{sub:integrating_non_terrestrial_layers}

\gls{geo} satellites, with their coverage umbrella of around 7,300 km of radius,  can provide stable and cost-effective connection for broadcast services. 
 However, the GEO channel suffers from a tremendous propagation delay of around 240 ms and severe path loss due to atmospheric absorption over great distances~\cite{cianca2005integrated}, which make point-to-point GEO deployments (Configuration GE in Fig.~\ref{fig:1}) not desirable for 5G/6G applications.
 Collaboration among aerial/space platforms has thus been proposed to improve network performance. While LAP/UAV-assisted networks are foreseen (e.g., to enhance public safety and support traffic requirements in hot-spot areas~\cite{zhao2019uav}, as illustrated in Fig.~\ref{fig:usecase}), we focus on satellite- and HAP-based deployments, which do not pose power/mobility constraints and are becoming a hot research topic.
 \smallskip

 \textbf{GEO-LEO Integration (Configuration GLE in Fig.~\ref{fig:1}).}
 The introduction of frequent and affordable orbital insertions in the \acrlong{leo} (typically between 600 and 1,500 km) opens up new opportunities for communication. 
 Compared to standalone GEO constellations, the GLE framework improves optical sensors resolution and geographical position accuracy thanks to the increased proximity to the targets, thus offering the same sensing efficiency with lighter payloads and smaller size or, equivalently, incorporating more capable platforms at the same cost. Moreover, it guarantees improved payload performance since LEO satellites, which operate in rarified environments, can implement aerodynamic deorbiting and maneuvers via air breathing electric propulsion. Furthermore, GLE  achieves better wireless coverage and lower latency since LEOs can amplify and relay the upstream GEO signals towards the ground. GLE also provides both dense and comprehensive coverage when GEO's ability to support connectivity over very large areas while being continuously visible from terrestrial terminals and LEO's flexibility are leveraged together.

 However, the high-speed mobility and related Doppler shift experienced in the lower orbits  require dense constellations of satellites to maintain signal continuity on Earth. A standalone space network may also complicate delay-sensitive delivery of data as the channel capacity has to be split among a very large number of ground terminals, thus saturating the available bandwidth.
  \smallskip



 \textbf{GEO-HAP Integration (Configuration GHE in Fig.~\ref{fig:1}).}
HAPs, which operate in the stratosphere, can act as wireless relays to improve global connectivity, as acknowledged in the literature~\cite{ye2020space}.
 Compared to the GE and GLE configurations, GHE
 	 not only improves capacity by amplifying the GEO signal before~forwarding it to the ground, but also ensures quicker and cheaper deployment. Also, it  offers adaptive networking capabilities as the network topology can be adjusted on demand based on instantaneous temporal and traffic demands. The GHE approach
 	 guarantees continuous end-to-end coverage as HAPs, unlike LEO satellites, operate in a quasi-stationary position, and allow communication equipment to operate with less interference and/or distortion. 
 	Additionally, GHE permits to host computing and storage facilities on HAPs, i.e., closer to the ground users, rather than on satellites, thus promoting better latency and reliability for applications like mission offloading thanks to the more favorable link budget in the HAP-Earth link.
 	However, while standalone GEO deployments can offer wide geographical connectivity, an intermediate HAP will necessarily reduce the footprint shaped on the ground, and should operate in constellations/swarms to ensure seamless~coverage.
 \smallskip

 \textbf{GEO-LEO-HAP Integration (Configuration GLHE in Fig.~\ref{fig:1}).}
A three-hop integrated network can further enhance communication performance by building a seamless reconfigurable network environment that provides a much larger coverage than a classic terrestrial network~\cite{liu2018space}. 
The upper layer is a bi-dimensional satellite network, which may be organized in a mesh topology to create an overlay access backbone switching network. 
GEO-LEO integration can mitigate network congestion by cross-migrating traffic requests to/from the GEO and LEO layers, thus improving load balancing.
The bottom layer is the aerial network based on HAPs, which may connect together for a larger regional coverage. 
Specifically, the aerial layer may act as a relay for connections between terrestrial users and the higher satellite layers.
This approach provides an additional degree of robustness in case one aerial/space platform is damaged, as 	other layers can temporarily serve terrestrial traffic requests.
However, although the hybrid GLHE system is superior to the traditional two-tier systems, it is hard to achieve full deployment due to its high cost and management complexity, especially on the satellite layers.

\begin{table*}[t!]
\renewcommand{\arraystretch}{1.7}
\caption{System parameters for space, air, and ground architectures~\cite{38821,ITU_F.2439-0}.}
    \label{tab:params}
    \centering
    \resizebox{\textwidth}{!}{%
\begin{tabular}{|l|ccccccccc|c|c}
\hline
\multirow{3}{*}{Parameter} & \multicolumn{8}{c|}{Space } & Aerial & \multicolumn{2}{c|}{Terrestrial} \\ \cline{2-12}

& \multicolumn{4}{c|}{Downlink}                                                                       & \multicolumn{4}{c|}{Uplink}                                                                         & \multirow{2}{*}{HAP}                                                                 & \multicolumn{2}{c|}{\multirow{2}{*}{Base station}} \\ \cline{2-9}
                           & \multicolumn{2}{c}{GEO}   & \multicolumn{2}{c|}{LEO}                                                & \multicolumn{2}{c}{GEO}   & \multicolumn{2}{c|}{LEO}                                                &                                                                                      & \multicolumn{2}{c|}{}                    \\ \cline{1-1} \cline{1-12} 
Altitude ($h$) [km]                   & \multicolumn{2}{c}{35,786} & \multicolumn{2}{c|}{\{1200, 600\}} & \multicolumn{2}{c}{35,786} & \multicolumn{2}{c|}{\{1200, 600\}} & 20                                                                                   & \multicolumn{2}{c|}{0.03}                   \\ \cline{1-1} \cline{11-12} \cline{2-10}
Frequency ($f_c$) [GHz]                 & 2            & 20         & 2                        & \multicolumn{1}{c|}{20}                      & 2           & 30          & 2                        & \multicolumn{1}{c|}{30}                      & 38                                                                                   & 2            & \multicolumn{1}{c|}{20}            \\ \cline{1-1} \cline{11-12} \cline{2-10}
Max. EIRP$^{\star}$ [dBW]                      & 73.8         & 66         & 54                       & \multicolumn{1}{c|}{36}                      & 73.8           & 46.2           & 48.6                     & \multicolumn{1}{c|}{46.2}                    & 27.9                                                                                 & N/A$^{\ddagger}$                 & \multicolumn{1}{c|}{N/A$^{\ddagger}$}                    \\ \cline{1-1}
System bandwidth  ($B$) [MHz]                         & 30           & 400        & 30                       & \multicolumn{1}{c|}{400}                     & 30          & 400         & 30                       & \multicolumn{1}{c|}{400}                     & 400                                                                                  & N/A$^{\ddagger}$                 & \multicolumn{1}{c|}{N/A$^{\ddagger}$}                     \\ \cline{1-1}
Rx. antenna-gain-to-noise-temperature$^{\dagger}$ ($G/T$) [dB/K]              & $-$31.6        & 15.9       & $-$31.6                    & \multicolumn{1}{c|}{15.9}                    &19          & 28           & 1.1                      & \multicolumn{1}{c|}{13}                      & 27.7                                                                                 & N/A$^{\ddagger}$     & \multicolumn{1}{c|}{N/A$^{\ddagger}$}       \\ \cline{1-10}
Rx. antenna gain ($G_R$) [dBi]             & \multicolumn{9}{c|}{N/A$^{\ddagger}$ (already included in $G/T$)}  & 0 & \multicolumn{1}{c|}{39.7}       \\ \cline{1-1}

Noise figure (NF) [dB]                        & \multicolumn{9}{c|}{N/A$^{\ddagger}$ (already included in $G/T$)}                                                                                                                                                                                                                                                                 & 7                 & \multicolumn{1}{c|}{1.2}                  \\ \cline{1-1}
Antenna temperature $T_a$ (K)       & \multicolumn{9}{c|}{N/A$^{\ddagger}$ (already included in $G/T$)}                                                                                                                                                                                                                                                                 & 290               & \multicolumn{1}{c|}{150}                  \\ \cline{1-1} \cline{11-12}
Ambient temperature $T_0$  (K)      & \multicolumn{9}{c|}{N/A$^{\ddagger}$ (already included in $G/T$)}                                                                                                                                                                                                                                                                 & \multicolumn{2}{c|}{290}                 \\ \cline{1-12} 
Fading                    & \multicolumn{8}{c|}{\begin{tabular}[c]{@{}c@{}}Shadowed-Rician  \\ $\{b_0,m,\omega\}=\{0.158,19.4,1.29\}$\cite{abdi2003new}\end{tabular}}                                                          & \begin{tabular}[c]{@{}c@{}}Rician  \\ $C = 10$\cite{alsharoa2020improvement}\end{tabular} & \multicolumn{2}{c|}{N/A$^{\ddagger}$}                   \\ \hline
 \multicolumn{12}{l}{$^{\ddagger}$Not available in the 3GPP specifications.\vspace{-0.1cm}}\\
 \multicolumn{12}{l}{$^{\star}$The EIRP accounts for the antenna transmit power, the cable loss, and the transmit antenna gain~\cite{38821}.\vspace{-0.1cm}}\\
\multicolumn{12}{l}{$^{\dagger}$$G/T$ accounts for the receive antenna gain, the ambient/antenna temperature, and the noise figure~\cite{38821}.}
\end{tabular}
}
\end{table*}

\section{System Design}
\label{sec:design}
In this section, we present the channel characterization for space/air communications according to the latest 3GPP and ITU guidelines  (Sec.~\ref{sub:channel_models}), and related system parameters (Sec.~\ref{sub:system_parameters}).

\subsection{Channel Models} 
\label{sub:channel_models}
We consider a downlink system model in which GEOs, LEOs, HAPs and terrestrial base stations form a 3D multi-hop network. Intermediate nodes adopt a fully cooperative amplify-and-forward (AF) relay protocol.
Considering a communication system in which a GEO signal propagates through $N$ hops before arriving to its destination, the \gls{snr} $\gamma^{(n)}_{i,j}$, $i,j\in\{\text{G,\,L,\,H,\,E}\}$, experienced at the $n-$th hop between transmitter $i$ and receiver $j$, is computed as
\begin{equation}
\label{eq:snr}
\gamma^{(n)}_{i,j} =  \text{EIRP}_i  + \frac{G_{j}}{T}  - \text{PL}_{i,j} + \tau_{i,j} - k - B - \text{NF}. \quad \rm{[dB]}
\end{equation}
In Eq.~\eqref{eq:snr}, EIRP$_i$ is the effective isotropic radiated power (which accounts for the antenna transmit power, the cable loss, and the transmit antenna gain~\cite{38821}), $G_j/T$ is the receive antenna-gain-to-noise-temperature (which accounts for the receive antenna gain, the ambient/antenna temperature, and the noise figure), PL is the path loss, $\tau$ is the fading, $k$ is the Boltzmann constant, $B$ is the channel bandwidth, and NF is the noise figure.
The end-to-end \gls{snr} for the complete AF system can then be expressed as 
\begin{equation}
	\gamma_{\rm AF}=\left[\prod_{n=1}^N\left(1+\frac{1}{\gamma_{ij}^{(n)}}\right)-1\right]^{-1}.
	\label{eq:e2e-snr}
\end{equation}

\subsubsection{Path Loss} 
\label{ssub:path_loss}
Besides free-space path loss, which increases with the carrier frequency and the propagation distance between the transmitter and the receiver, in a fully integrated space-air-ground framework the signal undergoes several other stages of attenuation due to:
\begin{itemize}
 	\item \emph{Atmospheric Gases $A_g$}, consisting of dry air  and water vapour attenuation.
 	\item \emph{Scintillation  $A_s$}, due to fluctuations of the refractive index in the ionosphere (atmospheric turbulence scintillation in the troposphere) for below (above) 3 GHz transmissions.
 \item \emph{Clutter Loss $L_c$}, due to signal absorption and diffraction by ground objects, such as buildings or vegetation.
 \item \emph{Rain Absorption $A_{r}$}, causing unavailability of signals due to scattering from rainfall (especially above 10 GHz).
\end{itemize}

It must be mentioned that, for space channels that do not involve complete penetration through the atmosphere (i.e., in the GEO-LEO, GEO-HAP, and LEO-HAP links), a simple free-space path loss model can be considered.
For a more complete description of the channel model in NTN scenarios, we refer the interested reader to~\cite{3GPP38811,giordani2020satellite}.


\subsubsection{Channel Fading}
For the space-to-ground link, Shadowed-Rician distribution has been proposed in~\cite{abdi2003new} to describe the large-scale fading due to amplitude fluctuation of the transmitted signal's envelope in the rarefied environment. 
In turn, a Rician distribution has been demonstrated to characterize more accurately  smaller-scale fading in the air-to-ground environment~\cite{alsharoa2020improvement}.
As expected, space-to-space and space-to-air channels should not be affected by fading, as the density of the air in these cases is much smaller than near the ground, so many of the fading effects, such as rainfall, atmosphere, and scattering, can be neglected.

\subsection{System Parameters} 
\label{sub:system_parameters}
System design parameters are summarized in Table~\ref{tab:params} for both satellite- and HAP-based architectures.
\smallskip 

\subsubsection{Space Design} 
According to the 3GPP specifications~\cite{38821}, LEO stations can be deployed at an altitude of 600 or 1200 km.
Satellites can operate both in the S-bands at 2 GHz or in the Ka-bands (i.e., within the \gls{mmwave} spectrum) at 20 (30) GHz for downlink (uplink) transmissions, with a bandwidth of 30 and 400 MHz, respectively.
 It should be noticed that, while GEO satellites can support direct transmission links on the ground, thanks to their very directional and high-power transmission (up to 73.8
dBW of EIRP in the S-bands) and large phased antennas offering fine electronic beam-steering (with a receive gain of up to 28 dB/K in the Ka-bands), LEO orbits are typically populated by nano/picosatellites incorporating simple electronic devices to reduce component costs, and are constrained by limited power and antenna gains.
\smallskip

\subsubsection{Aerial Design} 
\label{ssub:hap_design_}
According to the ITU guidelines~\cite{ITU_F.2439-0}, HAPs operate in the Ka-bands at 38 GHz with a bandwidth of 400 MHz.
While guaranteeing an ultra-flexible deployment, these elements offer an antenna-gain-to-noise-temperature of up to 27.7 dB/K without the prohibitive costs of satellite infrastructures.
\smallskip 

\subsubsection{Terrestrial Design}
For \gls{mmwave}-enabled terrestrial base stations, a receiving antenna offering a gain of 39.7 dBi should be adopted~\cite{38821}, while omnidirectional unit gain should be considered at sub-6 GHz.

\section{Multi-Layered NTNs: \\Performance Comparison}
\label{sec:results}
In this section we numerically compare the performance of the different multi-layered NTN configurations introduced in Sec.~\ref{sec:multi}.
The optimal architecture is then identified by evaluating (i) the average ergodic (Shannon) capacity, and (ii) the outage probability, i.e., the probability that the received signal on the Earth is below a predefined threshold $\epsilon$. In AF systems, assuming that all system parameters are independently distributed, the outage probability $\Phi_{\rm AF}$ can be calculated as
\begin{equation}
\Phi_{\rm AF} = 1-\left(\prod_{n=1}^N \mathbb{P}\left[\gamma^{(n)}_{i,j}>\epsilon\right]\right), \: \forall\, i,\,j\in\{\text{G,\,L,\,H,\,E}\}.
\end{equation}

Our analysis, which is based on the channel model~described in Sec.~\ref{sub:channel_models}, has been validated by Monte Carlo simulations obtained generating 10\,000 random realizations of the fading for each investigated integrated NTN configuration.
Results are given as a function of the carrier frequency $f_c$, the signal quality threshold $\epsilon$ (varied from $-$20 to 40 dB to consider different levels of sensitivity at the receiver), the LEO altitude~$h_L$ (which only affects those configurations in which LEO satellites are deployed, i.e., GLE and GHLE), and the elevation angle $\alpha$ between the transmitter and the receiver. Air/spaceborne stations are deployed on the same line, even though an additional average pointing loss of 0.5 dB is considered to characterize the loss in signal strength due to misalignment between the transmitting and receiving antennas at every hop.
 \smallskip




\begin{figure}[t!]
\centering
  \begin{subfigure}[t!]{0.44\textwidth}
  \centering
    \includegraphics[width=1\columnwidth]{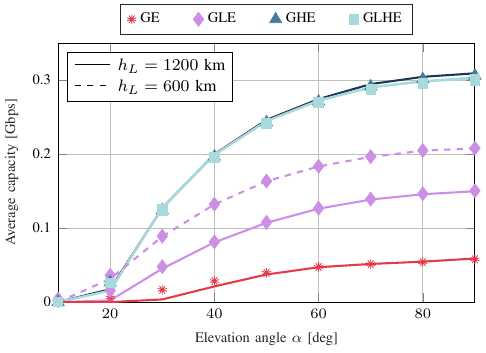}
    \caption{$f_c=2$ GHz.}
      \label{fig:capacity-2}
  \end{subfigure} \\\vspace{0.33cm}
  \begin{subfigure}[t!]{0.44\textwidth}
  \centering
    \includegraphics[width=1\columnwidth]{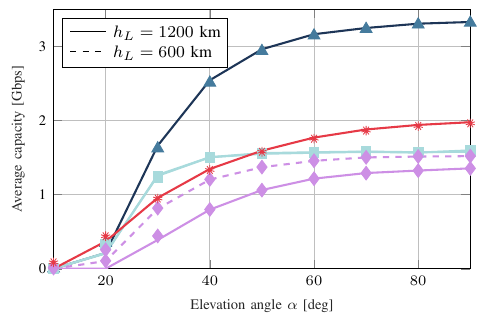}
    \caption{$f_c=20 \: (30)$ GHz for DL (UL).} 
      \label{fig:capacity-20}
  \end{subfigure}
  \setlength{\belowcaptionskip}{-0.33cm}
\caption{Average ergodic capacity vs. the elevation angle, as a function of the operational frequency for different multi-layered NTN architectures, when $\epsilon=0$ dB. The lines represent the results from the theoretical model, and the markers indicate Monte Carlo simulations.}
	  \label{fig:capacity} 
\end{figure}

\textbf{Average Capacity.}
In Fig.~\ref{fig:capacity} we compare the average capacity achieved by different NTN configurations.
First, we observe that the numerical results closely follow the theoretical curves obtained applying Shannon's formula for the channel capacity, which is related to the end-to-end \gls{snr} expressed in Eq.~\eqref{eq:e2e-snr} and the available bandwidth, thereby validating our framework.
Second, the very large bandwidth available in the Ka-bands (Fig.~\ref{fig:capacity-20}) delivers 10$\times$ higher capacity than in the S-bands (Fig.~\ref{fig:capacity-2}), which are constrained by the limited frequency resources available in the 2-GHz  channel.
Moreover,  despite the worse propagation conditions experienced at high frequency, the physical size of the antennas is so small that it becomes practical to install large antenna arrays on non-terrestrial platforms, thereby scaling up network performance by beamforming (e.g., for GEO satellites, the antenna-gain-to-noise-temperature $G/T$ is increased from $-$31.6 to 15.9 dB/K at 20 GHz).

As expected, at 2 GHz the standalone GE configuration shows the worst capacity performance due to the fact that atmospheric perturbations at very large distances severely deteriorate the signal's propagation without proper countermeasures. Conversely, all investigated multi-layered architectures guarantee better coverage by amplifying GEO's transmission before forwarding it to the ground. In particular, GHE offers up to 6$\times$ higher capacity than GE and emerges as the most preferable integration choice for NTNs. A fully integrated GLHE scheme achieves comparable performance but, due to the three dimensions and the mobility of the space layer, is much more complex than its GHE counterpart.

 \begin{figure}[t!]
\centering
  \begin{subfigure}[t!]{0.44\textwidth}
  \centering
    \includegraphics[width=1\columnwidth]{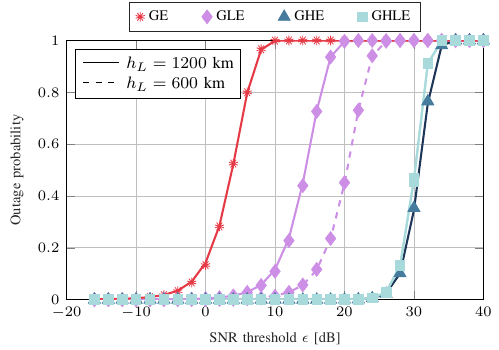}
    \caption{$f_c=2$ GHz.}
      \label{fig:outage-2}
  \end{subfigure} \\\vspace{0.15cm}
  \begin{subfigure}[t!]{0.44\textwidth}
  \centering
    \includegraphics[width=1\columnwidth]{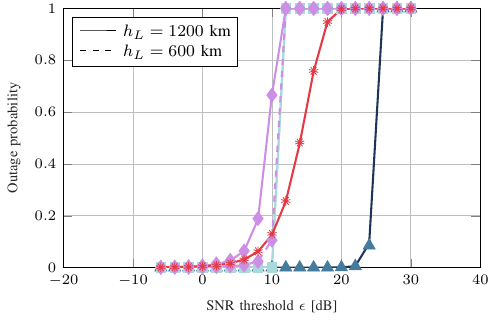}
    \caption{$f_c=20 \: (30)$ GHz for DL (UL).}  
      \label{fig:outage-20}
  \end{subfigure}
  \setlength{\belowcaptionskip}{-0.33cm}
\caption{Outage probability vs. SNR threshold, as a function of the operational frequency for different  multi-layered NTN architectures, when $\alpha=90$ deg. The lines represent the results from the theoretical model, and the markers indicate Monte Carlo simulations.}
	  \label{fig:outage} 
\end{figure}

Furthermore, better capacity can be provided when a standalone GEO layer is assisted by  HAPs operating in the stratosphere (in GHE) rather than by  LEO satellites  (in GLE). In fact, the former approach allows to decrease the length of the (bottleneck) Earth link traversing the atmosphere to only 20 km, i.e., where HAPs are typically deployed, compared to the 600 or 1200 km of the latter, thus resulting in more robust channels. 
In any case, while 3GPP specifications~\cite{38821} support two deployment options for LEOs, Fig.~\ref{fig:capacity} demonstrates that it is more desirable to form a LEO constellation at $h_L=600$ km to benefit from the resulting lower path loss in the shorter space-Earth link. 
This, in turn, does not affect GLHE's design (which shows perfectly overlapping curves for $h_L=600$ and $1200$ km) since LEO-HAP propagation is nearly in free-space, where a difference of 600 km in the propagation distance accounts only for a (negligible) 6 dB in the overall link budget. 

When transitioning to the Ka-bands, Fig.~\ref{fig:capacity-20} shows that LEO-based architectures like GLE and GLHE now are outperformed even by the baseline GE configuration due to a simpler hardware implementation in LEO compared to GEO satellites: for example, Table~\ref{tab:params} indicates that LEO's EIRP for relaying operations is as low as 36 dBW at 20 GHz, compared to the 66 dBW of GEO satellites.

Finally, Fig.~\ref{fig:capacity} illustrates that the system performance decreases at low elevation due to the more serious impact of clutter and atmospheric loss, as the signal has to transit longer through the atmosphere~\cite{giordani2020satellite}. 
This effect is particularly critical in the GHE and GLHE configurations where the visibility of HAPs, which fly at low altitude, is impaired for $\alpha<20$~deg.
\smallskip

 \begin{figure}[t!]
\centering
  \begin{subfigure}[t!]{0.44\textwidth}
  \centering
   \includegraphics[width=0.97\columnwidth]{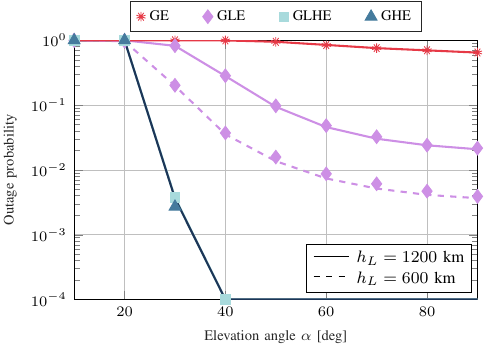}
    \caption{$f_c=2$ GHz.}
      \label{fig:outage-el-2}
  \end{subfigure} \\\vspace{0.15cm}
  \begin{subfigure}[t!]{0.44\textwidth}
  \centering
    \includegraphics[width=0.97\columnwidth]{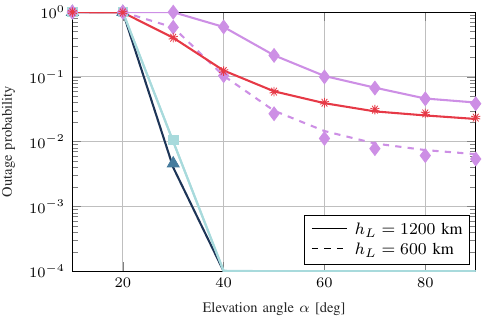}
    \caption{$f_c=20 \: (30)$ GHz for DL (UL).} 
      \label{fig:outage-el-20}
  \end{subfigure}
  \setlength{\belowcaptionskip}{-0.33cm}
\caption{Outage probability vs. the elevation angle, as a function of the operational frequency for different multi-layered NTN architectures, when $\epsilon=5$ dB. The lines represent the results from the theoretical model, and the markers indicate Monte Carlo simulations.}
	  \label{fig:outage-el} 
\end{figure}

\textbf{Outage Probability.}
Similar conclusions can be derived from Fig.~\ref{fig:outage}, which reports the outage probability $\Phi_{\rm AF}$ for different values of the SNR threshold $\epsilon$ in various NTN configurations, in both S- and Ka-bands. 
We observe that $\Phi_{\rm AF}$, which is an indication of the quality of the communication, decreases when decreasing $\epsilon$, i.e., considering more complex (and expensive) receivers able to detect the signal in more noisy channels. 
Moreover, we see that integrating multiple non-terrestrial layers together offers better coverage compared to a standalone design: at 2 GHz an outage probability lower that $1\%$ (a critical requirement for most 5G/6G-oriented services) is achieved for $\epsilon=24$ dB in the GHE structure, against $\epsilon=-6$ dB in the GE case, thus ensuring more robust and continuous connectivity at lower costs.
This is because the intermediate HAP permits to establish shorter-range communications in the Earth link, thus ensuring better signal quality and stronger received power.
As already mentioned, LEO relays exhibit lower performance gains as channel conditions in the space layer are already sufficiently good to guarantee satisfactory signal quality even without amplification.

Furthermore, we observe that NTNs operating in the S-bands (Fig.~\ref{fig:outage-2}) provide more reliable communications than in the Ka-bands (Fig.~\ref{fig:outage-20}) since they present a lower outage probability. This is due to the increased variability of the channel conditions at 20/30 GHz due to clutter loss and reflection from terrain and buildings, and the more severe impact of atmospheric absorption. 
Specifically, LEO-based configurations experience the most significant performance deterioration by more than 20 dB compared to communications in the S-bands: this is due to, besides worse propagation characteristics, the simpler hardware design in LEO satellites.
The multi-layered  GHE architecture, although suffering, on average, from a 5 dB performance loss, still stands out at the most desirable option to minimize the outage probability.

\begin{figure}[t!] 
 \centering
 \includegraphics[width=0.95\columnwidth]{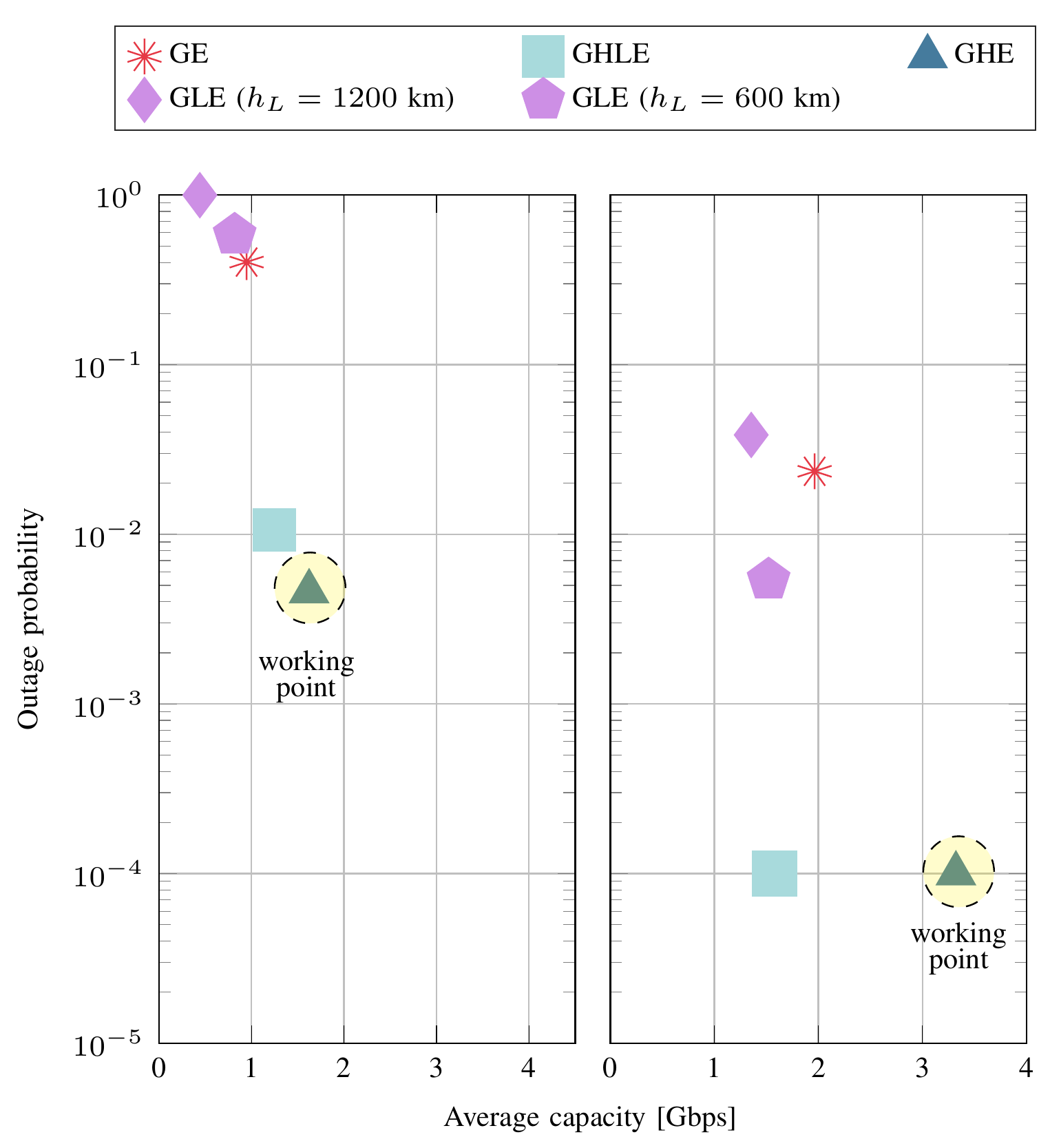}
      \caption{Outage probability vs. average ergodic capacity trade off for different multi-layered NTN architectures. We set $\alpha=30$ deg (left), $\alpha=90$ deg (right), $\epsilon=5$ dB, and $f_c=20 \: (30)$ GHz for DL (UL).}
      \label{fig:comparison}
\end{figure}%

Finally, Fig.~\ref{fig:outage-el} illustrates that the outage performance improves consistently when increasing the elevation angle~$\alpha$, as a result of the shorter path of the signal and the lower impact of scintillation.
In this case, it appears evident that point-to-point GE deployments are certainly not compatible with robustness. 
In particular, the long transmission distance in the GEO-Earth link makes $\Phi_{\rm AF}$ increase above 50\% for all values of $\alpha$, even when considering perfect angular alignment (i.e., $\alpha=90$ deg). 
Outage is also unacceptably high for $\alpha<20$ deg in all investigated configurations, even though communication might still be established in the open (rural) environment where clutter loss is negligible.
In any case, the multi-layered approach permits to support more reliable  communications (with values of outage probability lower than $10^{-4}$) by relaying the GEO signal which would otherwise be  undetectable.
It should also be noticed that, while GE's  performance improves when operating at 20 GHz because of the larger antenna gains achievable by beamforming, LEO relaying can guarantee lower outage probability only when intermediate satellites are deployed at $h_L=600$ km, i.e., when the endpoints of the space-Earth link are progressively closer. 
\smallskip 

\textbf{Comparison.} 
To summarize the conclusions from the previous paragraphs, Fig.~\ref{fig:comparison} compares the outage vs. capacity performance of different multi-layered hierarchical networks. As anticipated, the deployment of intermediate HAPs in the stratosphere (configuration GHE) ensures up to $1.75\times$ better capacity in the Ka-bands than point-to-point GEO transmissions (configuration GE), while resulting in more robust communications. The GHE approach outperforms even the GLE configuration by 2 orders of magnitude in~terms of outage probability and by more than $2\times$ in terms of~average capacity, thus making it clear that LEO satellites are not desirable to relay the upstream signal towards the ground.  
Furthermore, a fully integrated space-air-ground scenario (configuration GLHE) does not enhance the system performance beyond GHE's (the GLHE capacity decreases by $42\%$ compared to GHE), while in turn resulting in more complex and expensive network management. Finally, Fig.~\ref{fig:comparison} (right) shows that better performance can be generally guaranteed at high elevation thanks to the shorter path and lower attenuation of the signal.



\section{Concluding Remarks}
\label{sec:conclusions}
6G research is just in its infancy and there remain many open challenges to solve, including whether and how to design NTNs to assist terrestrial communication.
In this work we addressed this issue by proposing multi-layered hierarchical networks in which the merits of the  space, air, and ground layers are incorporated together to improve quality of service. Specifically, we compared the performance of different cooperative architectures against a standalone GEO constellation, and evaluated which degree of integration offers better capacity and outage probability.
Our results proved that HAP relays (configuration GHE) can best bridge the satellite signal to the ground while ensuring up to 6$\times$ better capacity than point-to-point GEO transmissions. We also demonstrated that hardware constraints in the Ka-bands make LEO-based relaying configurations not desirable.


As part of our future work, we will consider end-to-end simulations to assess the benefits of the proposed multi-layered integrations in terms of network-related metrics such as the overall transmission latency, the achievable throughput, and the packet delivery rate. Moreover, we will analyze the effect of a dynamic network scenario, thereby accounting for the intrinsic mobility of LEO and HAP relays.



\begin{IEEEbiography}
[{\includegraphics[width=0.9in,clip,keepaspectratio]{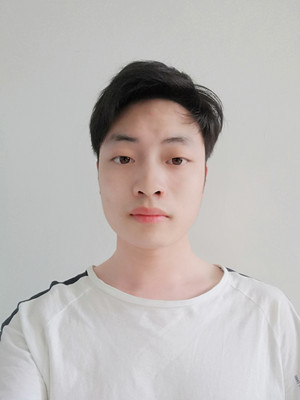}}]{Dengke Wang}
 was born in Sichuan, China. He received the B.Sc. degree from the University of Electronic Science and Technology of China in 2019. Now, he is pursuing the M.Sc. degree in King Abdullah University of Science and Technology (KAUST), Saudi Arabia. His main research interests include performance analysis and modeling of wireless communication networks and software-defined radio.
\end{IEEEbiography}%

\begin{IEEEbiography}
[{\includegraphics[width=0.9in,clip,keepaspectratio]{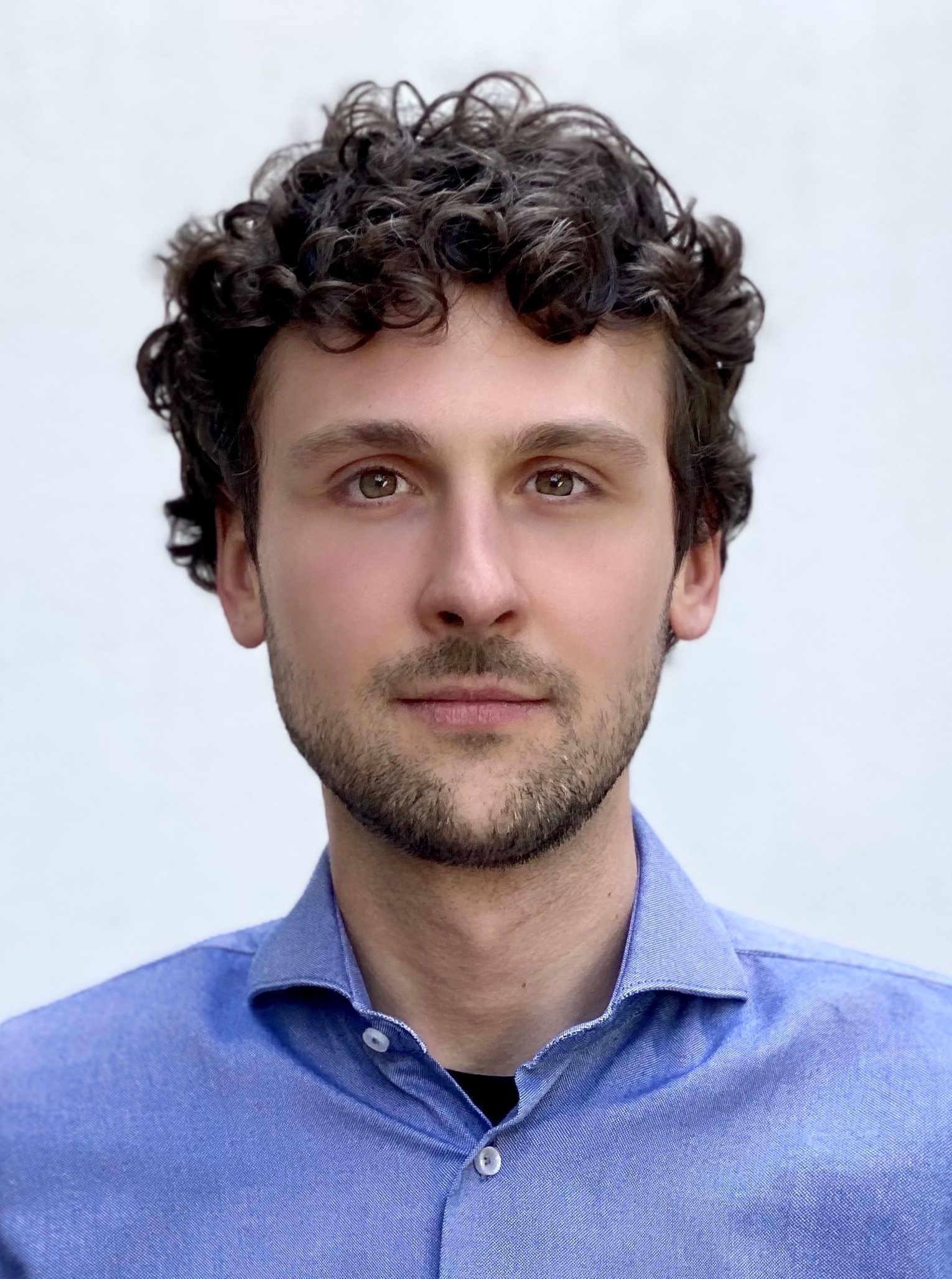}}]{Marco Giordani}
[M'20] received his Ph.D. in Information Engineering in 2020 from the University of Padova, Italy,  where he is now a postdoctoral researcher and adjunct professor.
He visited  NYU and TOYOTA Infotechnology Center, Inc., USA.
In 2018 he received the “Daniel E. Noble Fellowship Award” from the IEEE Vehicular Technology Society. His research  focuses on protocol design for 5G/6G mmWave cellular and vehicular networks.
\end{IEEEbiography}%

\begin{IEEEbiography}
[{\includegraphics[width=0.9in,clip,keepaspectratio]{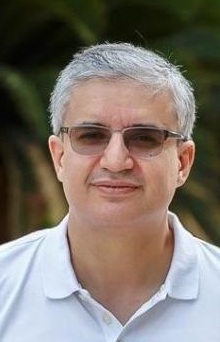}}]{Mohamed-Slim Alouini}
[S'94-M'98-SM'03-F'09] was
born in Tunis, Tunisia. He received the Ph.D.
degree in Electrical Engineering from the
California Institute of Technology (Caltech),
Pasadena, CA, USA, in 1998. He served as a faculty
member in the University of Minnesota, Minneapolis,
MN, USA, then in the Texas A\&M University at Qatar,
Education City, Doha, Qatar before joining KAUST,
Thuwal, Makkah Province, Saudi Arabia as a
Professor of Electrical Engineering in 2009. His
current research interests include the modeling,
design, and performance analysis of wireless
communication systems.
\end{IEEEbiography}%

\begin{IEEEbiography}
[{\includegraphics[width=0.9in,clip,keepaspectratio]{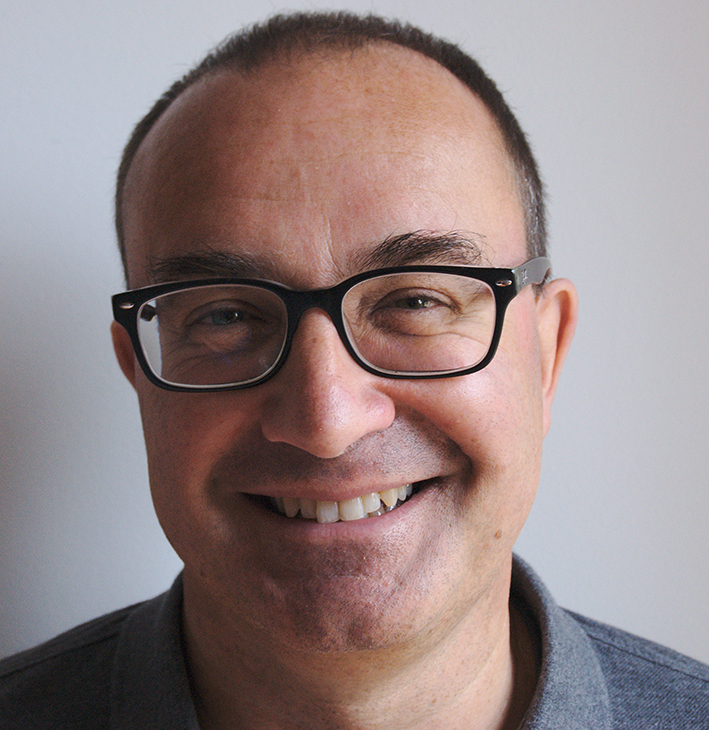}}]{Michele Zorzi}
[F'07] is with the Information Engineering Department of the University of Padova, focusing on wireless communications research. He was Editor-in-Chief of IEEE Wireless Communications from 2003 to 2005, IEEE Transactions on Communications from 2008 to 2011, and IEEE Transactions on Cognitive Communications and Networking from 2014 to 2018. He served ComSoc as a Member-at-Large of the Board of Governors from 2009 to 2011, as Director of Education and Training from 2014 to 2015, and as Director of Journals from 2020 to 2021.
\end{IEEEbiography}


\end{document}